\documentclass{article}%
\usepackage{amsmath}
\usepackage{amssymb}
\usepackage{hyperref}
\usepackage{amsfonts}
\usepackage{graphicx}%
\begin{document}

\title{On Enceladian Fields}
\author{T.S. Van Kortryk$^{\ddag}$, T.L. Curtright$^{\S \dag}$, and H.
Alshal$^{\S \Delta\sharp}\medskip$\\$^{\S }$Department of Physics, University of Miami, Coral Gables, Florida 33124\\$^{\bigtriangleup}$Department of Physics, Faculty of Science, Cairo
University, Giza, 12613, Egypt\\$^{\ddag}${\footnotesize vankortryk@gmail.com\ \ \ \ \ }$^{\dag}%
${\footnotesize curtright@miami.edu\ \ \ \ \ }$^{\sharp}$%
{\footnotesize halshal@sci.cu.edu.eg\medskip}}
\date{\textit{In tribute to
\href{https://en.wikipedia.org/wiki/Murray_Gell-Mann}{Murray Gell-Mann}
(1929-2019).}}
\maketitle

\begin{abstract}
We fantasize about an alternative history for theories of electromagnetism and gravitation.

\end{abstract}

\begin{quote}
\textquotedblleft In the field of Astrobiology, the precise location,
prevalence and age of potential extraterrestrial intelligence have not been
explicitly explored.\textquotedblright\ --- Cai et al. \cite{astrobio}
\end{quote}

\vfill

\href{https://en.wikipedia.org/wiki/List_of_adjectivals_and_demonyms_of_astronomical_bodies}{The
Enceladians} have developed an advanced society, an achievement quite
remarkable for water-dwelling,
\href{https://en.wikipedia.org/wiki/Bioluminescence}{bioluminescent} creatures
whose physical features are reminiscent of
\href{https://en.wikipedia.org/wiki/Cephalopod#Distribution_and_habitat}{the
Cephalopoda} living at great depths in terrestrial seas \ (But they live much
longer than terrestrial cephalopoda --- some as many as 500 Earth years!)
\ Their scientific knowledge is impressive, although founded on experiments
limited in spatial extent to the domain of their existence, an enclosed sea
sheathed in a shell of ice many kilometers deep that surrounds their
\href{https://en.wikipedia.org/wiki/Enceladus}{small Saturnian moon}. \ They
have no direct knowledge of anything outside this frozen crust, having never
tried to pierce through it or even to probe its full thickness, mostly because
of religious beliefs that have served them well for the past several
millennia. Nevertheless, they have developed theories of electromagnetism, and
of gravity.

They communicate mostly through images and light, particularly in one-on-one
\textquotedblleft conversations\textquotedblright\ by modifying the
luminescent patterns on their skin, but also through the use of sound.
\ Consequently, they understand very well the classical wave nature of both
light and sound, as might be expected of highly intelligent beings with such
ample life spans. \ Also, given their aquatic, ice-bound world, with the
recurring deformations imposed on it by their unseen host planet, it is not
surprising that they have long understood the mathematics of tensor calculus,
especially in the context of continuum mechanics.

And so it happens, based on considerable thought and very careful experiments
within the confines of their world, the Enceladians have ascertained that in
almost all situations sound can be accounted for by a scalar field, $\phi$,
while electromagnetism can be explained using a vector field, $\psi_{\mu}$,
but to encode all its subtleties gravity is best described by a symmetric
tensor, $\chi_{\mu\nu}$. On these points we are in agreement.\footnote{But of
course, the Enceladians do \emph{not} use Greek letters. \ We have
transliterated their nomenclature.}

However, unlike us, they are not enamored with the concept of gauge invariance
or the beauty of zero mass. \ They view zero mass as a peculiar limit,
probably pathological, and the corresponding invariances as most likely just
symptoms of some poorly understood illness for the underlying theory. \ Thus
they describe classical electromagnetism using a Klein-Gordon field equation
\cite{Notation} with vector current sources, $J_{\mu}$.
\ Explicitly,\footnote{We take the liberty to use relativistic notation. (The
Enceladians do not!) \ We use $\eta_{\mu\nu}=\operatorname{diag}\left(
+1,-1,-1,-1\right)  =\eta^{\mu\nu}$ to raise and lower indices, and sum over
the same if repeated. \ }%
\begin{equation}
\left(  \square+m^{2}\right)  \psi_{\mu}=J_{\mu}+\frac{1}{m^{2}}~\partial
_{\mu}\partial_{\nu}J^{\nu} \label{Spin1FE}%
\end{equation}
Here $J_{\mu}$ is most likely but not necessarily conserved, and for the
Enceladians the value of $m$ amounts to a question to be answered by
experiment. \ So far their experiments only set limits on $m$ and on the
conservation of $J_{\mu}$. \ Nonetheless, they understand that even if
$\partial_{\nu}J^{\nu}\neq0$, so long as $m\neq0$ the divergence
$\partial^{\mu}\psi_{\mu}$ does not correspond to any scalar radiation given
off by the source.

Similarly, most Enceladian scientists describe gravity using the field
equation%
\begin{equation}
\left(  \square+M^{2}\right)  \chi_{\mu\nu}=T_{\mu\nu}+\dfrac{1}{M^{2}}\left(
\partial_{\mu}\partial^{\rho}T_{\rho\nu}+\partial_{\nu}\partial^{\rho}%
T_{\rho\mu}-\frac{1}{2}~\eta_{\mu\nu}\partial^{\rho}\partial^{\sigma}%
T_{\rho\sigma}\right)  +\frac{2}{3M^{4}}\left(  \partial_{\mu}\partial_{\nu
}-\frac{1}{4}~\eta_{\mu\nu}\square\right)  \partial^{\rho}\partial^{\sigma
}T_{\rho\sigma} \label{Spin2FE}%
\end{equation}
where $T_{\mu\nu}$ is a symmetric, traceless (i.e. $0\equiv\eta_{\mu\nu}%
T^{\mu\nu}$) tensor source, not necessarily conserved, and where the value of
$M$ is also to be determined by experiment. \ So far their experiments again
only set limits on $M$ and on the conservation of $T_{\mu\nu}$. \ And again,
they are aware that even if $\partial^{\rho}T_{\rho\nu}\neq0$, so long as
$M\neq0$ the divergence $\partial^{\mu}\chi_{\mu\nu}$ does not correspond to
either vector or scalar radiation from localized $T_{\mu\nu}$'s, while the
trace $\chi=\eta_{\mu\nu}\chi^{\mu\nu}$\ is also not produced as scalar
radiation from such sources. \ Indeed, directly from (\ref{Spin2FE}), $\left(
\square+M^{2}\right)  \chi=0$ for traceless $T_{\mu\nu}$, so $\chi$ is a free
field of mass $M$.\footnote{The Enceladians are also aware that their method
can be extended to higher spins without limit. \ For any $J$, with the higher
spin field described by a totally symmetric rank $J$ tensor, there are source
terms involving as many as $2J$ derivatives of a traceless, symmetric, rank
$J$ source tensor on the RHS of the field equation, whose form is a
straightforward generalization of (\ref{Spin2FE}). \ It is not necessary for
that source tensor to be conserved.}

Ongoing experimental studies \cite{EJP} in their laboratories have convinced
some Enceladians that the tensor $T_{\mu\nu}$ is not conserved, in general,
but in fact its divergence can always be written as just the gradient of a
scalar, $S$. \ Therefore, the full tensor source may be written in terms of a
conserved tensor, $\Theta_{\mu\nu}$, plus this scalar part, $T_{\mu\nu}%
=\Theta_{\mu\nu}+\eta_{\mu\nu}S$. \ The scalar is then constrained by the
traceless condition on $T_{\mu\nu}$ to be $S=-\Theta/4$, where $\Theta
=\eta_{\mu\nu}\Theta^{\mu\nu}$. \ Thus,%
\begin{equation}
T_{\mu\nu}=\Theta_{\mu\nu}-\frac{1}{4}~\eta_{\mu\nu}\Theta\ ,\ \ \ \partial
^{\rho}\Theta_{\rho\nu}=0\ ,\ \ \ \partial^{\rho}T_{\rho\nu}=-\frac{1}%
{4}~\partial_{\nu}\Theta
\end{equation}
So written, their gravitational field equation becomes%
\begin{equation}
\left(  \square+M^{2}\right)  \chi_{\mu\nu}=\Theta_{\mu\nu}-\frac{1}{4}%
~\eta_{\mu\nu}\Theta-\dfrac{1}{6M^{4}}\left(  \partial_{\mu}\partial_{\nu
}-\frac{1}{4}~\eta_{\mu\nu}\square\right)  \left(  \square+3M^{2}\right)
\Theta\label{AnotherSpin2FE}%
\end{equation}
Consequently, not only is the trace $\chi$ a free field of mass $M$, but the
divergence $\partial^{\mu}\chi_{\mu\nu}$ decouples as well, in the following
sense.%
\begin{equation}
\left(  \square+M^{2}\right)  \left[  \partial^{\mu}\chi_{\mu\nu}+\dfrac
{1}{8M^{4}}\left(  \square+2M^{2}\right)  \partial_{\nu}\Theta\right]  =0
\end{equation}
That is to say, the combination within the brackets $\left[  \cdots\right]  $
is also a free field of mass $M$. \ 

Recently, one of the more careful thinkers among Enceladian physicists (whose
actual \textquotedblleft name\textquotedblright\ is usually displayed
bioluminescently on its skin, in a pattern most easily remembered by humans as
the acronym \textquotedblleft MGM\textquotedblright) has realized a change of
dependent variable that gives an alternative expression for their theory of
gravity. \ Namely, after making the substitution%
\begin{equation}
\chi_{\mu\nu}\rightarrow\chi_{\mu\nu}+\frac{1}{24M^{4}}\eta_{\mu\nu}\left(
2M^{2}+\square\right)  \Theta-\frac{1}{6M^{4}}\partial_{\mu}\partial_{\nu
}\Theta
\end{equation}
MGM obtained a simpler, indeed elegant, field equation%
\begin{equation}
\left(  \square+M^{2}\right)  \chi_{\mu\nu}=\Theta_{\mu\nu}-\frac{1}{3}%
~\eta_{\mu\nu}\Theta-\frac{1}{3M^{2}}~\partial_{\mu}\partial_{\nu}%
\Theta\label{NewSpin2FE}%
\end{equation}
and subsequently referred to it as a \textquotedblleft
teeter-totter\textquotedblright\ partial differential
equation\footnote{\textquotedblleft Teeter-totter\textquotedblright\ is of
course our translation of MGM's Enceladian description, and is not to be
confused with \href{https://en.wikipedia.org/wiki/Seesaw_mechanism}{the
seesaw\ mechanism} familiar to terrestrial physicists.}: \ As the $M^{2}$ term
on the LHS decreases, the $1/M^{2}$ term among the RHS sources increases, and
vice versa. \ In any case, it follows from (\ref{NewSpin2FE}) that the trace
and the divergence of $\chi_{\mu\nu}$ decouple on a more equal
footing\footnote{For the Enceladians \textquotedblleft
footing\textquotedblright\ should be understood in the
\href{https://en.wikipedia.org/wiki/Muscular_hydrostat}{muscular hydrostatic
sense}.} inasmuch as%
\begin{gather}
\left(  \square+M^{2}\right)  \left[  \chi+\tfrac{1}{3M^{2}}~\Theta\right]
=0\label{Trace}\\
\left(  \square+M^{2}\right)  \left[  \partial^{\mu}\chi_{\mu\nu}+\tfrac
{1}{3M^{2}}~\partial_{\nu}\Theta\right]  =0 \label{Divergence}%
\end{gather}
That is to say, the two combinations of fields and sources shown in
(\ref{Trace}) \& (\ref{Divergence}) are both free fields of mass $M$. \ 

After obtaining (\ref{NewSpin2FE}), MGM gave some consideration to its
phenomenological consequences. \ Naively, from (\ref{NewSpin2FE}), the mass
$M$ cannot be too large, or the exponential suppression of static fields would
be plainly evident even in experiments confined to the modest number of
kilometers accessible to the Enceladians. \ Nor can $M$ be too small if
$\Theta\neq0$, or else the $\partial_{\mu}\partial_{\nu}\Theta$ source on the
RHS of (\ref{NewSpin2FE}) would be enhanced by the $1/M^{2}$ factor to the
point of disagreement with various moderate precision, time-dependent
experiments. \ Similar statements about $m$ follow from (\ref{Spin1FE}) if
$\partial_{\nu}J^{\nu}\neq0$, but for theoretical as well as experimental
reasons, the Enceladians have set strong limits on $\partial_{\nu}J^{\nu}$ and
for most purposes treat $J_{\mu}$ as a conserved current. \ In that case,
their experiments lead directly to upper limits on $m$.

Because of the size limitations imposed by their habitat and the environment
in which they must work, direct experiments to set high precision limits on
$M$ are difficult but not impossible for the Enceladians. \ Looking for an
exponential height variation in static gravitational fields is the best they
can do, arriving at $M\leq3\times10^{-47}\ kg$ corresponding to a distance
scale of about ten kilometers. \ Similarly, electromagnetic wave dispersion
experiments over distances of several kilometers present many challenges for
them to infer a value for $m$, due mostly to background effects in their
watery world. \ Consequently, their best limits come from electrostatic and
magnetostatic experiments using meter-scale devices to find $m\leq
10^{-45}\ kg$. \ (For comparison to terrestrial results, see \cite{PhotonMass}
and \cite{GravitonMass}.) \ 

However, as MGM first emphasized, the limits on $m$ and $M$ are sufficiently
different that they pose a potential problem. \ If the actual non-zero values
of $m$ and $M$ are close to the Enceladian limits, then the ratio
$m/M\approx30$ which would imply the last\ source term in (\ref{NewSpin2FE})
is easily the most significant, and relatively quite large for many processes
that can be measured with precision and without competing backgrounds, for
example processes where light is effected by gravity. \ For smaller values of
$M$, the problem becomes more pronounced, perhaps to the point of being in
conflict with experiments on the scattering of light by light. \ In detail,
the Enceladian expression of $\Theta_{\mu\nu}$ for electromagnetic fields in
the absence of any $J_{\mu}$ sources is%
\begin{equation}
\Theta_{\mu\nu}=-\kappa\left[  \partial_{\mu}\psi^{\sigma}~\partial_{\nu}%
\psi_{\sigma}-\frac{1}{2}\eta_{\mu\nu}\left(  \partial^{\rho}\psi^{\sigma
}~\partial_{\rho}\psi_{\sigma}-m^{2}\psi^{\lambda}\psi_{\lambda}\right)
+\dfrac{1}{6}\left(  \eta_{\mu\nu}\square-\partial_{\mu}\partial_{\nu}\right)
\psi^{\lambda}\psi_{\lambda}\right]
\end{equation}
where $\kappa$ is a constant determined in static experiments involving masses
on a torsion balance.\footnote{Indeed, within their experimental uncertainties
the Enceladians do find $\kappa=16\pi G/c^{4}$, i.e. $G=6.67\times10^{-11}%
%TCIMACRO{\unit{m}}%
%BeginExpansion
\operatorname{m}%
%EndExpansion
^{3}%
%TCIMACRO{\unit{kg}}%
%BeginExpansion
\operatorname{kg}%
%EndExpansion
^{-1}%
%TCIMACRO{\unit{s}}%
%BeginExpansion
\operatorname{s}%
%EndExpansion
^{-2}$ after conversion to terrestrial SI units.} \ This tensor source is
conserved, given (\ref{Spin1FE}) with $J_{\mu}=0$, and has a non-zero trace.%
\begin{gather}
\partial^{\mu}\Theta_{\mu\nu}=-\kappa\left[  \partial_{\nu}\psi_{\sigma
}~\left(  \square+m^{2}\right)  \psi^{\sigma}\right]  \circeq0\\
\Theta=-\kappa\left[  m^{2}\psi^{\lambda}\psi_{\lambda}+\psi_{\sigma}\left(
\square+m^{2}\right)  \psi^{\sigma}\right]  \circeq-\kappa m^{2}\psi^{\lambda
}\psi_{\lambda}%
\end{gather}
Here \textquotedblleft$\circeq$\textquotedblright\ means equality given
$0=\left(  \square+m^{2}\right)  \psi^{\sigma}$. \ So the $1/M^{2}$ term in
(\ref{NewSpin2FE}) is%
\begin{equation}
\frac{\kappa m^{2}}{3M^{2}}~\partial_{\mu}\partial_{\nu}\left(  \psi^{\lambda
}\psi_{\lambda}\right)  \label{Trouble}%
\end{equation}
Generically, this is $10^{3}$ larger than the $\Theta_{\mu\nu}$ term on the
RHS of (\ref{NewSpin2FE}), upon evaluating $m/M$ right at the Enceladian
limits given previously. \ This is experimentally untenable in many
situations, even given the limitations of Enceladian science. \ MGM's
conclusion is that whatever the direct experimental limits on $M$ might be, if
(\ref{Trouble}) is less than or comparable to experimentally constrained
effects of the $\Theta_{\mu\nu}$ source, then $m$ cannot be much larger than
$M$. \ 

But what if the Enceladian theory of gravity is flawed? \ Indeed, MGM has
realized that a large $m/M$ coefficient is probably a signal that additional
contributions have been neglected in the source for $\chi_{\mu\nu}$. \ For
large $m/M$ most likely there are nonlinear terms involving the $\chi$-field
coupled to itself. \ MGM is grappling with nonlinear extensions of the theory,
but has not yet managed to solve the problem. \ In that respect, MGM is
considerably behind some mid-20$^{th}$ century terrestrial developments
\cite{OP,PGOF}.

Nevertheless, MGM's point about unusually large coupling of electromagnetic
fields to gravity, for small but non-zero $M$, should be taken seriously by
terrestrial physicists. \ The present value of $m/M$ taken from the Particle
Data Group summaries \cite{PhotonMass,GravitonMass}, if evaluated right at the
best accepted experimental limits, would give $m/M\approx2\times10^{13} $~?!
\ So either $m$ is actually much smaller than the current experimental limits,
and/or $M$ is much larger, or else higher-order nonlinear effects that have
been omitted from the Enceladian model, so far, must be included to make it
realistic. \ 

Unknown to the Enceladians, even if the photon is absolutely massless, there
is still an anomalous quantum trace for the electromagnetic energy-momentum
tensor \cite{CE}, namely, $\Theta=-\kappa N\hbar\alpha F^{\rho\sigma}%
F_{\rho\sigma}$ with $F_{\mu\nu}$ the usual electromagnetic field strength,
$\alpha$ the fine-structure constant, and $N$ a model-dependent number
proportional to $\sum_{j}q_{j}^{2}\left(  -1\right)  ^{2s_{j}}\left(
1-12s_{j}^{2}\right)  $ where the sum is over all physical particle states
with helicities $s_{j}$ and electric charges $q_{j}e$. \ In the Enceladian
framework associated with field equation (\ref{NewSpin2FE}) this would lead to
a source term enhanced by $1/M^{2}$ as given by%
\begin{equation}
\frac{\kappa N\hbar\alpha}{M^{2}}~\partial_{\mu}\partial_{\nu}\left(
F^{\rho\sigma}F_{\rho\sigma}\right)  \label{MoreTrouble}%
\end{equation}
For electric fields of frequency $\omega$ this term could give generic
contributions to $\chi_{00}$ of order $\kappa\alpha\hbar^{2}\omega^{2}%
E^{2}/M^{2}c^{4}$, thereby dominating the leading $\Theta_{\mu\nu}$ source
contribution of order $\kappa E^{2}$, if
\begin{equation}
Mc^{2}\ll\hbar\omega\sqrt{\alpha} \label{DisparityCondition}%
\end{equation}
For frequencies as low as one MHz, this is troublesome since in that case
$\left.  \hbar\omega\sqrt{\alpha}\right\vert _{\omega=2\pi\times10^{6}}%
\approx4\times10^{-10}~eV$, while the previously mentioned, rather crude
Enceladian limit gives $Mc^{2}<2\times10^{-11}~eV$. \ This would give an
enhancement of (\ref{MoreTrouble}) compared to $\Theta_{\mu\nu}$ by two orders
of magnitude! \ For higher $\omega$ the situation is worse. \ So here again,
higher-order nonlinear effects that have been omitted from the Enceladian
model, so far, would have to be included to make it realistic.

The situation brings to mind Vainshtein's proposal \cite{V,BD} that nonlinear
gravitational effects can be exploited to avoid the van
Dam-Veltman-Zakharov-Iwasaki\ discontinuity \cite{VDV}, where the latter was
based on single particle exchange diagrams for massive versus massless
gravitons. \ Only in that case, the disparity between experiments (planetary
orbits compared to bending of light by the sun) caused by the discontinuity
was merely a factor of $3/4=O\left(  1\right)  $ whereas a disparity expressed
by (\ref{DisparityCondition}) could easily be many orders of magnitude. \ 

The Enceladians had better get to work on those nonlinear terms!\bigskip

\noindent\textbf{Acknowledgements:} \ Supported in part by a University of
Miami Cooper Fellowship, this work was written in honor of
\href{https://en.wikipedia.org/wiki/Murray_Gell-Mann}{Murray Gell-Mann}.
\ While Gell-Mann's thoughts on gravity, supergravity, and superstrings are
\href{https://inspirehep.net/literature?sort=mostrecent&size=25&page=1&q=FIND%20AU%20gell-mann%20and%20ti%20supergravity}{well-documented
in the literature}, along with his fascination for
\href{https://inspirehep.net/literature?sort=mostrecent&size=25&page=1&q=FIND%20AU%20gell-mann%20and%20title%20histories}{alternative
histories}, evidence for his thinking about science fiction is largely
\href{https://cgc.physics.miami.edu/MGMShort.pdf}{anecdotal}, but extremely
plausible given the breadth of his interests going all the way back to his
youth.\footnote{\textquotedblleft But even then I did have some thoughts about
the future of the human race, especially in connection with the textbooks and
the scientific romances of
\href{https://en.wikipedia.org/wiki/H._G._Wells}{H.G. Wells}. \ I loved to
read his novels ... \textquotedblright\ --- p 14,
\textit{\href{https://www.amazon.com/s?k=the+quark+and+the+jaguar}{\textit{The
Quark and the Jaguar}}}}

\subsection*{Appendix}

A one-parameter field equation that leads to pure spin 2 radiation from a
conserved source $\Theta_{\mu\nu}$ is given by%
\begin{equation}
\left(  \square+M^{2}\right)  \chi_{\mu\nu}=\Theta_{\mu\nu}+\frac{r-1}{4}%
~\eta_{\mu\nu}\Theta+\frac{r}{3M^{2}}\left(  \eta_{\mu\nu}\square
-\partial_{\mu}\partial_{\nu}\right)  \Theta+\frac{1-r}{24M^{4}}\left(
\eta_{\mu\nu}\square-4\partial_{\mu}\partial_{\nu}\right)  \left(
\square+3M^{2}\right)  \Theta\tag{A1}\label{Spin2FEr}%
\end{equation}
Choosing the parameter $r=0$ or $r=1$ gives simpler forms: \ The original O-P
equation \cite{OP} for spin 2 is obtained for $r=1$ while the
Enceladian\ equation (\ref{AnotherSpin2FE}) is given by $r=0$. \ Moreover, the
substitution%
\begin{equation}
\chi_{\mu\nu}\rightarrow\chi_{\mu\nu}+\frac{1}{24M^{4}}\eta_{\mu\nu}\left(
\left(  2+6r\right)  M^{2}+\left(  1-r\right)  \square\right)  \Theta+\frac
{1}{6M^{4}}\left(  r-1\right)  \partial_{\mu}\partial_{\nu}\Theta\tag{A2}%
\end{equation}
amounts to a field redefinition that leads to the equation
\begin{equation}
\left(  \square+M^{2}\right)  \chi_{\mu\nu}=\Theta_{\mu\nu}-\frac{1}{24M^{4}%
}\left(  \left(  8+6r\right)  M^{4}+5M^{2}r\square-r\square^{2}\right)
\eta_{\mu\nu}\Theta-\frac{1}{6M^{4}}\left(  \left(  2+r\right)  M^{2}%
+r\square\right)  \partial_{\mu}\partial_{\nu}\Theta\tag{A3}%
\end{equation}
This greatly simplifies upon making the choice $r=0$, to obtain the form found
by MGM.%
\begin{equation}
\left(  \square+M^{2}\right)  \chi_{\mu\nu}=\Theta_{\mu\nu}-\frac{1}{3}%
~\eta_{\mu\nu}\Theta-\frac{1}{3M^{2}}~\partial_{\mu}\partial_{\nu}%
\Theta\tag{A4}\label{A4}%
\end{equation}
It should be kept in mind that all choices for $r$ lead to acceptable
classical theories for the radiation of pure spin 2 fields of mass $M$. \ On
the other hand, if the relative coefficients on the RHS\ of (\ref{Spin2FEr})
are modified, the spin 2 radiation may well be accompanied by spin 1 and/or
spin 0 radiation \cite{OP,PGOF}. \ In particular, to introduce spin zero
radiation in the form of a propagating trace, $\chi$, while retaining the
decoupled divergence, in the sense of (\ref{Divergence}), it is only necessary
to change both factors of $1/3$ on the RHS of (\ref{A4}) to some other common
value, say, $s/3$.

Dual formulations of massive gravity are also interesting, and may lead to
additional understanding of the theory. \ For field equations this is easily
done. \ To convert the symmetric tensor pure spin 2 $\chi_{\mu\nu}$ theory to
a corresponding $\tau_{\left[  \lambda\mu\right]  \nu}$ tensor theory, or to
convert a mixed spin 2 and spin 0 theory into a corresponding $\tau
_{\lambda\mu\nu}$ tensor theory, where now $\tau_{\lambda\mu\nu}=\tau_{\left[
\lambda\mu\right]  \nu}\oplus\tau_{\left[  \lambda\mu\nu\right]  }$, it is
sufficient to act with $\varepsilon^{\alpha\beta\gamma\delta}\partial
^{\lambda}$ on both the field and the source terms in the $\chi_{\mu\nu}$
field equation, followed by contraction and symmetrization of the various
indices. \ See \cite{CA}.


\begin{thebibliography}{99}                                                                                               %


\bibitem {astrobio}Xiang Cai, Jonathan H. Jiang, Kristen A. Fahy, and Yuk L.
Yung, \textquotedblleft A Statistical Estimation of the Occurrence of
Extraterrestrial Intelligence in the Milky Way Galaxy\textquotedblright%
\ \href{https://arxiv.org/abs/2012.07902}{https://arxiv.org/abs/2012.07902}

\bibitem {Notation}For the Enceladians, $m$ and $M$ in the field equations are
really inverse length scales, $1/\ell$ and $1/L$. \ We have expressed the
equations in terms of masses through the familiar (to us) Compton wavelength
relation, e.g. $L=\hbar/\left(  Mc\right)  $. \ But the Enceladians have not
yet discovered $\hbar$, or anything else about quantum mechanics for that matter.

\bibitem {EJP}Enceladian Journal of Physics, \textit{to appear}.

\bibitem {PhotonMass}In addition to
\href{http://pdg.lbl.gov/2019/listings/rpp2019-list-photon.pdf}{the PDG photon
summary}, which gives the limit $mc^{2}<1\times10^{-18}~eV$, i.e.
$m<2\times10^{-54}~kg$, the history of experiments to determine $m$ is
reviewed by Liang-Cheng Tu, Jun Luo, and George T. Gillies, \textquotedblleft
The mass of the photon\textquotedblright%
\ \href{https://iopscience.iop.org/article/10.1088/0034-4885/68/1/R02}{Rep.
Prog. Phys. 68 (2005) 77--130}. \ As noted by Tu et al., terrestrial scale
experiments on the dispersion of radio waves in the 1930-40s gave
$m\leq5\times10^{-46}~kg$. \ (See Table 1 for more recent experiments.)
\ Maxwell's 1870s redo of Cavendish's experiment was not as good, by an order
of magnitude, $m\leq5\times10^{-45}~kg$. \ (See Table 2 for more recent
experiments.) \ Of course, all the current extra-terrestrial methods (e.g. as
summarized in Table 3 of Tu et al.) are \emph{not} \emph{yet} available to the Enceladians.

\bibitem {GravitonMass}%
\href{http://pdg.lbl.gov/2019/listings/rpp2019-list-graviton.pdf}{The PDG
graviton summary} gives the best experimental limit as $Mc^{2}<6\times
10^{-32}~eV$, i.e. $M\lessapprox1\times10^{-67}~kg$.

\bibitem {OP}V.I. Ogievetsky and I.V. Polubarinov, \textquotedblleft
Interacting field of spin 2 and the Einstein equations\textquotedblright%
\ \href{https://doi.org/10.1016/0003-4916(65)90077-1}{Ann.Phys. 35 (1965)
167--208}.

\bibitem {PGOF}P.G.O. Freund, A. Maheshwari, and E. Schonberg,
\textquotedblleft Finite Range Gravitation\textquotedblright%
\ \href{http://adsabs.harvard.edu/doi/10.1086/150118}{Astro.Journal 157 (1969)
857--867}.

\bibitem {CE}M.S. Chanowitz and J. Ellis, \textquotedblleft Canonical Trace
Anomalies\textquotedblright%
\ \href{https://doi-org.access.library.miami.edu/10.1103/PhysRevD.7.2490}{Phys.Rev.
D (1973) 2490-2506}.

\bibitem {V}A.I. Vainshtein, \textquotedblleft To the problem of nonvanishing
gravitation mass\textquotedblright%
\ \href{https://doi.org/10.1016/0370-2693(72)90147-5}{Phys. Lett. B39 (1972)
393-394}.

\bibitem {BD}E. Babichev, C. Deffayet, \textquotedblleft An introduction to
the Vainshtein mechanism\textquotedblright%
\ \href{https://iopscience.iop.org/article/10.1088/0264-9381/30/18/184001}{Class.
Quantum Grav. 30 (2013) 184001},
\href{https://arxiv.org/abs/1304.7240}{arXiv:1304.7240 [gr-qc]}

\bibitem {VDV}H. van Dam and M.J.G. Veltman, \textquotedblleft Massive and
mass-less Yang-Mills and gravitational fields\textquotedblright%
\ \href{https://doi.org/10.1016/0550-3213(70)90416-5}{Nucl. Phys. B22 (1970)
397-411}; V.I. Zakharov, \textquotedblleft Linearized Gravitation Theory and
the Graviton Mass\textquotedblright%
\ \href{http://www.jetpletters.ac.ru/ps/1734/article_26353.shtml}{JETP Lett.
12 (1970) 312}; Y. Iwasaki, \textquotedblleft Consistency Condition for
Propagators\textquotedblright%
\ \href{https://doi.org/10.1103/PhysRevD.2.2255}{Phys. Rev. D2 (1970) 2255}.

\bibitem {CA}T.L. Curtright and H. Alshal, \textquotedblleft Massive Dual Spin
2 Revisited\textquotedblright%
\ \href{https://doi.org/10.1016/j.nuclphysb.2019.114777}{Nucl. Phys. B 948
(2019) 114777},\ \href{https://arxiv.org/abs/1907.11532}{arXiv:1907.11532
[hep-th]}; \ H. Alshal and T.L. Curtright,\textquotedblleft Massive Dual
Gravity in N Spacetime Dimensions\textquotedblright%
\ \href{https://doi.org/10.1007/JHEP09(2019)063}{JHEP 09 (2019) 063},
\href{https://arxiv.org/abs/1907.11537}{arXiv:1907.11537 [hep-th]}
\end{thebibliography}
\end{document}